\documentstyle[12pt]{article}

\def\be{\begin{equation}}
\def\ee{\end{equation}}
\def\bea{\begin{eqnarray}}
\def\eea{\end{eqnarray}}

\def\nnmb{\nonumber}

\def\ol{\overline}

\def\1{\={1}}
\def\AA{$\mid$}
\def\MM{$\rangle\mid$}
\def\ZZ{$\rangle\!\!$}

\begin{document}

\begin{titlepage}


\makebox{ }\hfill NUHEP-TH-99-75 \\ 
\makebox{ }\hfill December 1999 \\ 

\vskip 0.5in

\begin{center}

{\large \bf 
$E_6$ unification model building I :\\
Clebsch-Gordan coefficients of $27\otimes \ol{27}$
}

\vskip 0.4in

Gregory W.~Anderson 
and
Tom\'{a}\v{s} Bla\v{z}ek$^{*}$ 

\vskip 0.2in

    {\em Department of Physics and Astronomy, Northwestern University,}\\
    {\em 2145 Sheridan Road, Evanston, IL 60208, USA}\\
\vskip 0.05in

e-mail: {\em ganderson@nwu.edu, tb@nwu.edu}

\vskip 0.2in

\end{center}

\vskip 0.7in

\begin{abstract}

In an effort to develop tools for grand unified model building for
the Lie group $E_6$, in this paper we present 
the computation of the Clebsch-Gordan coefficients
for the product (100000) $\otimes$ (000010), where
(100000) is the fundamental 27-dimensional representation
of $E_6$ and (000010) is its charged conjugate.
The results are presented in terms of the dominant weight states of the
irreducible representations in this product.
These results are necessary for the group analysis 
of $E_6$ operators involving also higher representations, which is the
next step in this project.  In this paper we apply the results to the 
construction of the operator ${\bf 27}^3$.

\end{abstract}

\vskip 1.5in


\vskip 0.1in

$^*${\footnotesize On leave of absence from 
the Dept. of Theoretical Physics, Comenius Univ., Bratislava, Slovakia.
    }

\end{titlepage}

\renewcommand{\thepage}{\arabic{page}}
\setcounter{page}{1}

\section{Introduction}

Within the last thirty years it has become clear that the fundamental
properties of elementary particles may be explained by considering
symmetries larger than the explicit symmetry of the particle interactions. 
The concept of (spontaneously) broken symmetry thus became a cornerstone in our 
understanding of the basic properties of matter.
The logical extension of this principle has in turn generated interest in 
the investigation of symmetry groups larger than those found in 
the Standard Model (SM) of elementary particles. 
In particular, it is possible that the SM gauge group, $SU(3)\times SU(2) \times U(1)$,
is the broken relic of an enlarged symmetry group which only becomes manifest at higher energies. 
Examples of simple groups which are candidates for such a unified theory include
$SU(5)$~\cite{GG}, $SO(10)$~\cite{so10}, $SU(6)$~\cite{su6}, or $E(6)$~\cite{Ramond}.
Due to the enlarged symmetry, unification models based on these groups 
are potentially very predictive.  In addition to
unifying the three Standard Model gauge interactions, the most ambitious models
based on these symmetry groups seek to explain the observed quantum numbers of 
the low energy spectrum, understand the pattern of charged fermion and neutrino 
masses and mixing angles, and predict the strength of a variety of suppressed 
rare processes,  --- and do all 
that with just a few terms in the Lagrangian (or superpotential).

The group $E_6$ has long belonged to the most prominent candidates along this path of 
research \cite{Ramond,e6_models}. It contains the previously mentioned groups $SU(5)$, $SO(10)$, 
and $SU(6)$ as its subgroups. It allows chiral representations and its 
{\bf 27} dimensional fundamental representation can fit the fifteen known fermions
comprising one generation along with the right-handed neutrino 
(these states form the {\bf 16} of $SO(10)$~\footnote{For convenience we list the states in
terms of the representations of the familiar $SO(10)$ subgroup.}), 
the two (required by supersymmetry) Higgs doublets 
together with their colored counterparts (the {\bf 10} of $SO(10)$), 
and an SO(10) singlet. Thus the $E_6$ symmetry may impose constraints 
linking together charged fermion, neutrino and Higgs sectors of the SM,
a feature quite distinctive from theories based on lesser symmetry groups.
In addition, like $SO(10)$, the gauge anomalies are automatically canceled unlike 
the $SU(N)$ models.

While very attractive, the $E_6$ model building has not been extensively 
developed due to the mathematical complexities associated with a rank 6 exceptional group. 
In particular, the only Clebsch-Gordan coefficients which have been computed are those 
for products of two ${\bf 27}$s, 
or two  ${\bf \ol{27} }$s \cite{pat_E6}. 
The purpose of this series of papers is to continue the work along the line 
envisioned in refs. \cite{pat_E6} and \cite{pat_SU5} and 
present the results of the computation of the Clebsch-Gordan coefficients 
of various products of irreducible representations (irreps), necessary for a construction of complete $E_6$ models.
Our approach has been pragmatic: since these results provide
basic tools for unification model building we adopt 
a straightforward procedure and calculate the complete set
of states starting from the highest weight state of each irrep.

In this paper, we start with the decomposition of the product 
${\bf 27}\otimes{\bf \ol{27} }$, and as a direct application we relate 
our results to the operator {\bf 27}$^3$, which is the only tenable dimension four operator 
contributing to charged fermion masses.
In the follow-up papers \cite{AB234}, we will address the products
involving {\bf 78} and {\bf 351} irreps of $E_6$ and apply them
to other simple operators, which one needs for understanding
the symmetry breaking sector and/or the origin of fermion mass hierarchies.
In section 2, we present a basic theoretical background for the
calculation. Section 3 contains the Clebsch-Gordan coefficients for the dominant
weight states in the product ${\bf 27}\otimes{\bf \ol{27}}$. 
These results are then used in section 4
where a one-to-one correspondence between the states labeled by weights 
(in terms of Dynkin labels) 
and the SM gauge group states (in terms of fields carrying specific quantum numbers)
is established and 
a {\bf 27}$^3$ operator is decomposed into the sum of the 
SM gauge group interaction terms.

\section{Mathematical Preliminaries}

In the next two sections we primarily focus on the $E_6$ tensor product 
\be
 {\bf 27} \otimes {\bf \ol{27}} = {\bf 650} \oplus {\bf 78} \oplus {\bf 1}.
\label{eq:product}
\ee
or, equivalently, 
\be
 (100000) \otimes (000010) = (100010) \oplus (000001) \oplus (000000)
\label{eq:product2}
\ee
in terms of the highest weights of each irrep.
We note in passing that the numbering of the simple roots in $E_6$ as well as
other conventions we choose closely follow refs.\  \cite{Slansky_PhysRep} 
and \cite{pat_E6}. Also, note a conceptual difference between 
a {\em weight} $(w)$ and {\em weight state} $\mid\!\! w\rangle $.
The former is a set of six integer labels (Dynkin coordinates, throughout this paper) 
while the latter is a vector in the representation space.  This distinction is
important when we have degenerate weights {\it i.e.}, when there are multiple states 
with the same weight.

Next, we discuss our formalism in detail, since we will refer to it in the analyses of products
of larger irreps~\cite{AB234}.
The construction of the complete set of states in product (\ref{eq:product}) 
starts with ${\bf 650}$
and the decomposition of its highest weight state into the highest weight states
of ${\bf 27}$ and ${\bf \ol{27}}$,
\be
 \mid\!\! 100010\rangle  = \,\mid\!\! 100000\rangle \mid\!\! 000010\rangle.
\label{eq:HWS}
\ee
This state is a level $0$ state of ${\bf 650}$. 

In order to obtain the Clebsch-Gordan decomposition of the states at the next level, 
one of six lowering operators is applied to this equation. Lowering 
operators belong to the group generators outside the Cartan sub-algebra,
and their action on a state of weight $w=(w_1,w_2,w_3,w_4,w_5,w_6)$ from the weight system of
${\bf 27}$ or ${\bf \ol{27}}$ is given explicitly by 
\bea
E_{-\alpha_1} \mid\!\! w\rangle &=& 
                     N_{-\alpha,w}\;\;  \mid\!\! w_1-2, w_2+1, w_3  ,w_4  ,w_5  ,w_6 \rangle
                     \;\;\;\;\;\;\;\;\;\;\;\;  \;\;\; {\rm if}\; w_1>0, \nnmb \\
E_{-\alpha_2} \mid\!\! w\rangle &=& 
                     N_{-\alpha,w}\;\;  \mid\!\! w_1+1, w_2-2, w_3+1,w_4  ,w_5  ,w_6 \rangle
                     \;\;\;\;\;\;              \;\;\; {\rm if}\; w_2>0, \nnmb \\
E_{-\alpha_3} \mid\!\! w\rangle &=& 
                     N_{-\alpha,w}\;\;  \mid\!\! w_1  , w_2+1, w_3-2,w_4+1,w_5  ,w_6+1 \rangle
                                               \;\;\; {\rm if}\; w_3>0, 
                                                                       \label{eq:e- }      \\
E_{-\alpha_4} \mid\!\! w\rangle &=& 
                     N_{-\alpha,w}\;\;  \mid\!\! w_1  , w_2  , w_3+1,w_4-2,w_5+1,w_6 \rangle
                     \;\;\;\;\;\;              \;\;\; {\rm if}\; w_4>0, \nnmb \\
E_{-\alpha_5} \mid\!\! w\rangle &=& 
                     N_{-\alpha,w}\;\;  \mid\!\! w_1  , w_2  , w_3  ,w_4+1,w_5-2,w_6 \rangle
                     \;\;\;\;\;\;\;\;\;\;\;\;  \;\;\; {\rm if}\; w_5>0, \nnmb \\
E_{-\alpha_6} \mid\!\! w\rangle &=& 
                     N_{-\alpha,w}\;\;  \mid\!\! w_1  , w_2  , w_3+1,w_4  ,w_5  ,w_6-2 \rangle  
                     \;\;\;\;\;\;\;\;\;\;\;\;  \;\;\; {\rm if}\; w_6>0, \nnmb 
\eea
and
\be
E_{-\alpha_i} \mid\!\! w\rangle \; = \; 0  \;\;\;\;\;\;  \;\;\;  {\rm if}\; w_i\leq 0
                      \;\;\;\;\;\;  \;\;\;  {\rm for \; any}\;\; i=1,\ldots 6. 
\;\;\;\;\;\;  \;\;\;  \;\;\;\;\;\;  \;\;\;  \;\;\;\;\;\;  \;\;\;  \;\;\;  
\label{eq:e-0}
\ee
In our convention, the overall normalization factor 
\be
N_{-\alpha,w} = +1,
\ee
for any $E_{-\alpha_i}$, or any $\mid\!\! w\rangle$ provided the new state exists.
Note that if the result is non-zero, the new weight is
obtained from $(w)$ by subtraction of the corresponding simple root $\alpha_i$, 
which follows directly from the algebra of the group (see e.g., 
\cite{Slansky_PhysRep} or \cite{Dynkin})
\be
[H_n,E_{-\alpha_i}] \; = \; -\, (\alpha_i)_n \;  E_{-\alpha_i}.
\ee
$H_n$'s are the diagonal generators of the Cartan sub-algebra.

When lowering states of higher irreps, the newly obtained weight is again $(w - \alpha_i)$. 
However, when degeneracies are encountered, the change in the normalization can be non-trivial.
We follow the practice that the lowering of a higher irrep state is derived from lowering the irrep 
states it is built from. In simpler cases (for non degenerate weights, or for successive
lowerings through an entire multiplet of a particular $SU(2)$ subgroup)
the normalization factor can be expressed as
\be
     N_{-\alpha_i,w} = + \; [w_i + N_{-\alpha_i,w+\alpha_i}^2 ]^{1/2},
\label{eq:N2}
\ee
where it is assumed that $N_{-\alpha_i,w+\alpha_i} = 0$ if weight $(w)$ could not be
obtained from $(w+\alpha_i)$ at the previous level. Relation (\ref{eq:N2}) generalizes
(\ref{eq:e- }) and (\ref{eq:e-0}). It implies that states of the 
${\bf 650}$ and ${\bf 78}$ 
(unlike the states of the ${\bf 27}$ and ${\bf \ol{27}}$)
 can be lowered by $E_{-\alpha_i}$ even 
if the corresponding weight coordinate $w_i\leq 0$. 

As an example, two level 1 states are obtained from (\ref{eq:HWS}) 
by lowering with $E_{-\alpha_1}$ and $E_{-\alpha_5}$:
\bea
   E_{-\alpha_1} \mid\!\! 100010\rangle &=&    
  [E_{-\alpha_1} \mid\!\! 100000\rangle ]\, \mid\!\! 000010\rangle
             +   \mid\!\! 100000\rangle\,[E_{-\alpha_1} \mid\!\! 000010\rangle] \;\; = \nnmb \\
                                        &=& 
                 \mid\!\! \bar{1}10000\rangle \mid\!\! 000010\rangle + 0           \;\; = \nnmb \\
                                        &=& 
                 \mid\!\! \bar{1}10000\rangle \mid\!\! 000010\rangle                      \nnmb \\
   E_{-\alpha_5} \mid\!\! 100010\rangle &=&    
  [E_{-\alpha_5} \mid\!\! 100000\rangle ]\, \mid\!\! 000010\rangle 
             +   \mid\!\! 100000\rangle\,[E_{-\alpha_5} \mid\!\! 000010\rangle] \;\; = \nnmb \\
                                        &=& 
             0 + \mid\!\! 100000\rangle \mid\!\! 0001\bar{1}0\rangle           \;\; = \nnmb \\
                                        &=& 
                 \mid\!\! 100000\rangle \mid\!\! 0001\bar{1}0\rangle,                     \nnmb
\eea
where we use $\bar{x}\equiv -x$. 

A second example may be a sequence of two lowerings by $E_{-\alpha_1}$ of level 4 state 
$\mid\!\! 2\bar{1}0001 \rangle = {\mid\!\!100000\rangle}{\mid\!\!1\bar{1}0001\rangle}$. 
At level 5 we get 
$\: E_{-\alpha_1}\, \mid\!\! 2\bar{1}0001\rangle\:=\:\sqrt{2} \, \mid\!\! 000001_1\rangle$ 
when lowering the state on the left side. $\sqrt{2}$ follows
from $w_1 = 2$, see eq.(\ref{eq:N2}), 
and is consistent with obtaining a sum of two terms on the
right side. Proceeding to level 6 we again find the normalization from (\ref{eq:N2}),
$\: E_{-\alpha_1}\, \mid\!\! 000001_1\rangle =\sqrt{2} \, \mid\!\! \bar{2}10001 \rangle$, but
this time the $\sqrt{2}$ results from $N_{-\alpha_1, (2\bar{1}0001)}^{\:2}=2$.
At this level, factors of 2 cancel out leading to simple relation 
$\mid\!\! \bar{2}10001)= \mid\!\! \bar{1}10000\rangle \mid\!\! \bar{1}00001\rangle$. 

In general, relation (\ref{eq:N2}) is insufficient 
when degenerate weights are involved.
The {\bf 650}, for example, contains five degenerate, linearly independent 
states of (000001) weight. As in the second example above, we label them 
with a subscript corresponding to the last lowering used to derive the state. 
At issue is how to lower the weight state $\mid\!\! 000001_i\rangle$
with $E_{-\alpha_j}$, $i\neq j$. Clearly, this can be decided once we
know the decomposition into the states of the ${\bf 27}$ and ${\bf \ol{27}}$.
In particular, 
$E_{-\alpha_2} \mid\!\! 000001_1\rangle = +1/\sqrt{2} \mid\!\! 1\bar{2}1001\rangle$, while
$E_{-\alpha_j} \mid\!\! 000001_1\rangle = 0$, for $j=3,4,5$. 

In this way, one can obtain all 650 linearly independent states from the
highest weight state, eq.(\ref{eq:HWS}).
At level 5, however, one finds that the five (000001) states span
over a six-dimensional space. The extra linearly independent state of the same weight,
orthogonal to the subspace occupied by the previous five, is the highest 
weight state of the ${\bf 78}$. Lowering this state one recovers the
complete weight system of the ${\bf 78}$ irrep. The existence of an 
orthogonal weight subspace which provides for the highest weight state 
of another irrep is a general property of {\em dominant} weights;
the weights with all Dynkin coordinates non-negative.

\section{Clebsch-Gordan coefficients for $27\otimes \ol{27}$}

As we have just discussed, 
complete weight systems can be obtained from the highest weight state
of the highest irrep in the product. However, many states are going to be
decomposed into terms with the same coefficients and as a result the full table listing all 
Clebsch-Gordan coefficients (CGCs) would contain just a few distinct values.
For that reason it is not
necessary to list the decomposition of all linearly independent states in the weight system.
Instead, it is sufficient to provide CGCs just for the dominant weight states.

There are three dominant weights in the product ${\bf 27}\otimes {\bf \ol{27}}$,
corresponding to the highest weights of 
${\bf 650}$, ${\bf 78}$ and the singlet, eq.(\ref{eq:product2}).
In table \ref{t:paths} we show the lowering paths to the (000001) and (000000) 
dominant weight states 
of the ${\bf 650}$ and ${\bf 78}$. For instance, a path $12345$ is a shorthand
notation for the sequence of five lowering operators 
$E_{-\alpha_1}E_{-\alpha_2}E_{-\alpha_3}E_{-\alpha_4}E_{-\alpha_5}$ applied 
(from right to left) to the highest weight state.
Lowering paths in table \ref{t:paths} are, in general, not unique. 
Other paths may lead to the same weight states; e.g., in the {\bf 650} 
we get the same state $\mid$000001$_2\rangle$ following paths 23451, 21345, 
23145, and 23415. 12345, the path to $\mid$000001$_1\rangle$, is an example of 
a unique path. However, note that we cannot obtain a non-trivial linear combination of the
states in table \ref{t:paths} by following a different lowering path: the weight spaces
of the (000001) and (000000) weights in {\bf 650} contain exactly 5 and 20 different
states, respectively, and that matches their dimensionality. The same is true for
the six (000000) weight states in {\bf 78}. This is in sharp contrast to the 
higher irreps of $E_6$, as will be discussed in~\cite{AB234}.

Tables \ref{t:cgc000001} and \ref{t:cgc000000} 
contain the Clebsch-Gordan coefficients for the dominant weight states 
in ${\bf 27}\otimes {\bf \ol{27}}$, together with the decomposition of the singlet 
state, marked {\em S} for brevity. The numbering of the degenerate weights is consistent 
with table \ref{t:paths}, $0_i$ being equivalent to $\mid\!\! 000000_i\rangle$.
The last row in the tables shows the overall normalization of the state in the
respective column.

\protect
\begin{table}
\caption{ 
Lowering paths to dominant weights in (100000)$\otimes$(000010)
}
\label{t:paths}
\begin{tabular}{|c|c||c|c||c|c|}
\hline
\multicolumn{4}{|c||}{                  } & \multicolumn{2}{c|}{                  } \\
\multicolumn{4}{|c||}{\em (100010) irrep} & \multicolumn{2}{c|}{\em (000001) irrep} \\
\multicolumn{4}{|c||}{                  } & \multicolumn{2}{c|}{                  } \\
\hline
 Weight    & Lowering & Weight    & Lowering & Weight    & Lowering \\
 state     & path     & state     & path     & state     & path     \\
\hline                                   
    $\mid\!\! 000001_1     \rangle$&  12345            &
    $\mid\!\! 000000_{\: 1}\rangle$&  1234563421362345 & 
    $\mid\!\! 000000_1     \rangle$&  12364534236       \\  
    $\mid\!\! 000001_2     \rangle$&  23451            &
    $\mid\!\! 000000_{\: 2}\rangle$&  1436522336445321 & 
    $\mid\!\! 000000_2     \rangle$&  23645341236       \\  
    $\mid\!\! 000001_3     \rangle$&  34521            &
    $\mid\!\! 000000_{\: 3}\rangle$&  2345163421362345 & 
    $\mid\!\! 000000_3     \rangle$&  36452341236       \\  
    $\mid\!\! 000001_4     \rangle$&  45321            &
    $\mid\!\! 000000_{\: 4}\rangle$&  2345123466334521 & 
    $\mid\!\! 000000_4     \rangle$&  43652341236       \\  
    $\mid\!\! 000001_5     \rangle$&  54321            &
    $\mid\!\! 000000_{\: 5}\rangle$&  2451334266334521 & 
    $\mid\!\! 000000_5     \rangle$&  54362341236       \\  
 & &$\mid\!\! 000000_{\: 6}\rangle$&  3643542236112345 & 
    $\mid\!\! 000000_6     \rangle$&  63452341236       \\  
 & &$\mid\!\! 000000_{\: 7}\rangle$&  3645234512364321 & & \\  
 & &$\mid\!\! 000000_{\: 8}\rangle$&  3645236123445321 & & \\  
 & &$\mid\!\! 000000_{\: 9}\rangle$&  3164522336445321 & & \\  
 & &$\mid\!\! 000000_{10}  \rangle$&  4354231266334521 & & \\  
 & &$\mid\!\! 000000_{11}  \rangle$&  4352163452364321 & & \\  
 & &$\mid\!\! 000000_{12}  \rangle$&  5123644336223451 & & \\  
 & &$\mid\!\! 000000_{13}  \rangle$&  5236144336223451 & & \\  
 & &$\mid\!\! 000000_{14}  \rangle$&  5362144336223451 & & \\  
 & &$\mid\!\! 000000_{15}  \rangle$&  5432163452364321 & & \\  
 & &$\mid\!\! 000000_{16}  \rangle$&  6453342236112345 & & \\  
 & &$\mid\!\! 000000_{17}  \rangle$&  6532144336223451 & & \\  
 & &$\mid\!\! 000000_{18}  \rangle$&  6345236123445321 & & \\  
 & &$\mid\!\! 000000_{19}  \rangle$&  6134522336445321 & & \\  
 & &$\mid\!\! 000000_{20}  \rangle$&  6213324436554321 & & \\  
\hline
\end{tabular}
\end{table}

\protect
\begin{table}
\caption{ 
{\bf CG coefficients for (000001) dominant weight in (100000)$\otimes$(000010).}
The numbers in the last row indicate the overall denominator for the entries
in the respective column.}
\label{t:cgc000001}
\tiny
\begin{tabular}{|c|ccccc|c|}
\hline
\mbox{ } & \multicolumn{5}{|c|}{\mbox{ }}                     & {\mbox{ }} \\
\mbox{ } & \multicolumn{5}{|c|}{\normalsize \em (100010) } & {\normalsize \em (000001) } \\
\mbox{ } & \multicolumn{5}{|c|}{\mbox{ }}                     & {\mbox{ }} \\
\cline{2-7}
 & $\mid$000001$_1\rangle$& $\mid$000001$_2\rangle$& $\mid$000001$_3\rangle$& 
   $\mid$000001$_4\rangle$& $\mid$000001$_5\rangle$& $\mid$000001$  \rangle$ \\
\hline
    $\mid$100000$\rangle\mid$\={1}00001$\rangle$ &  1  &     &     &     &     &   1 \\

$\mid$\={1}10000$\rangle\mid$1\={1}0001$\rangle$ &  1  &  1  &     &     &     &  -1 \\

$\mid$0\={1}1000$\rangle\mid$01\={1}001$\rangle$ &     &  1  &  1  &     &     &   1 \\

$\mid$00\={1}101$\rangle\mid$001\={1}00$\rangle$ &     &     &  1  &  1  &     &  -1 \\

$\mid$000\={1}11$\rangle\mid$0001\={1}0$\rangle$ &     &     &     &  1  &  1  &   1 \\

    $\mid$0000\={1}1$\rangle\mid$000010$\rangle$ &     &     &     &     &  1  &  -1 \\
\hline
 & $\sqrt{2}$ & $\sqrt{2}$ & $\sqrt{2}$ & $\sqrt{2}$ & $\sqrt{2}$ & $\sqrt{6}$ \\
\hline
\end{tabular}
\end{table}

\protect
\begin{table}
\caption{ 
{\bf CG coefficients for (000000) dominant weight in (100000)$\otimes$(000010).}
The numbers in the last row indicate the overall denominator for the entries
in the respective column.
}
\label{t:cgc000000}
\tiny
\begin{tabular}{|@{\hspace{0.5mm}}c
                |c@{$\,$}c@{$\,$}c@{$\,$}c@{$\,$}c@{$\,$}c@{$\,$}c@{$\,$}c@{$\,$}c@{$\,$}c@{$\,$}
                 c@{$\,$}c@{$\,$}c@{$\,$}c@{$\,$}c@{$\,$}c@{$\,$}c@{$\,$}c@{$\,$}c@{$\,$}c@{$\,$}
                |c@{$\,$}c@{$\,$}c@{$\,$}c@{$\,$}c@{$\,$}c@{$\,$}
                |c@{$\,$}|}
\hline
\mbox{ } & &\multicolumn{19}{ c|}{\mbox{ }} & \multicolumn{6}{|c|}{\mbox{ }} & {\mbox{ }} \\
\mbox{ } & &\multicolumn{19}{ c|}{\normalsize \em (100010) } & 
                                              \multicolumn{6}{|c|}{ \normalsize \em (000001) } &
                                                                   {\normalsize \em S} \\
\mbox{ } & &\multicolumn{19}{ c|}{\mbox{ }} & \multicolumn{6}{|c|}{\mbox{ }} & {\mbox{ }} \\

\cline{2-28}
 & 
   $\,$0$_{1\:}$ & $\,$0$_{2\:}$ & $\,$0$_{3\:}$ & $\,$0$_{4\:}$ & $\,$0$_{5\:}$ & 
   $\,$0$_{6\:}$ & $\,$0$_{7\:}$ & $\,$0$_{8\:}$ & $\,$0$_{9\:}$ &     0$_{10} $ & 
   0$_{11}$ & 0$_{12}$ & 0$_{13}$ & 0$_{14}$ & 0$_{15}$ & 
   0$_{16}$ & 0$_{17}$ & 0$_{18}$ & 0$_{19}$ & 0$_{20}$ & 
   0$_1$ & 0$_2$ & 0$_3$ & 0$_4$ & 0$_5   $ &  0$_6   $ &  0 \\ 
\hline
\AA100000\MM\100000\ZZ    &  1&  &  &  &  &  &  &  &  &  &  &  &  &  &  &  &  &  &  &  & 1&  &  &  &  &  & 1 \\ 
\AA\110000\MM1\10000\ZZ   &  1&  & 1&  &  &  &  &  &  &  &  &  &  &  &  &  &  &  &  &  & 1& 1&  &  &  &  &-1 \\ 
\AA0\11000\MM01\1000\ZZ   &   &  & 1&  &  & 1&  &  &  &  &  &  &  &  &  &  &  &  &  &  &  & 1& 1&  &  &  & 1 \\ 
\AA00\1101\MM001\10\1\ZZ  &   &  &  &  &  & 1&  &  &  &  &  &  &  &  &  & 1&  &  &  &  &  &  & 1& 1&  & 1&-1 \\ 
\AA000\111\MM0001\1\1\ZZ  &   &  &  &  &  &  &  &  &  &  &  &  &  &  &  & 1& 1&  &  &  &  &  &  & 1& 1&-1& 1 \\ 
\AA00010\1\MM000\101\ZZ   &   &  &  &  &  &  &  &  &  & 1&  &  &  &  &  & 1&  &  &  &  &  &  &  &-1&  & 1& 1 \\
\AA0000\11\MM00001\1\ZZ   &   &  &  &  &  &  &  &  &  &  &  &  &  &  &  &  & 1& 1&  &  &  &  &  &  & 1& 1&-1 \\ 
\AA001\11\1\MM00\11\11\ZZ &   &  &  &  &  & 1&  &  &  & 1&  &  &  & 1&  & 1& 1&  &  &  &  &  &-1&-1&-1&-1&-1 \\
\AA0010\1\1\MM00\1011\ZZ  &   &  &  &  &  &  &  & 1&  &  &  &  &  & 1&  &  & 1& 1&  &  &  &  & 1&  &-1& 1& 1 \\ 
\AA01\1010\MM0\110\10\ZZ  &   &  & 1&  &  & 1&  &  &  &  &  &  & 1& 1&  &  &  &  &  &  &  &-1&-1&  & 1&  & 1 \\
\AA01\11\10\MM0\11\110\ZZ &   &  &  &  & 1&  &  & 1&  & 1&  &  & 1& 1&  &  &  &  &  &  &  & 1& 1& 1& 1&  &-1 \\ 
\AA1\10010\MM\1100\10\ZZ  &  1&  & 1&  &  &  &  &  &  &  &  & 1& 1&  &  &  &  &  &  &  &-1&-1&  &  &-1&  &-1 \\
\AA010\100\MM0\10100\ZZ   &   &  &  & 1& 1&  &  &  &  & 1&  &  &  &  &  &  &  &  &  &  &  &-1&  & 1&  &  & 1 \\ 
\AA1\101\10\MM\110\110\ZZ &   & 1&  &  & 1&  &  &  &  &  &  & 1& 1&  &  &  &  &  &  &  & 1& 1&  &-1&-1&  & 1 \\
\AA\100010\MM1000\10\ZZ   &  1&  &  &  &  &  &  &  &  &  &  & 1&  &  & 1&  &  &  &  &  &-1&  &  &  & 1&  & 1 \\
\AA1\11\100\MM\11\1100\ZZ &   & 1&  & 1& 1&  &  & 1& 1&  &  &  &  &  &  &  &  &  &  &  &-1&-1&-1&-1&  &  &-1 \\ 
\AA\1001\10\MM100\110\ZZ  &   & 1&  &  &  &  &  &  &  &  & 1& 1&  &  & 1&  &  &  &  &  & 1&  &  & 1& 1&  &-1 \\
\AA10\1001\MM\10100\1\ZZ  &   &  &  &  &  &  &  & 1& 1&  &  &  &  &  &  &  &  & 1& 1&  & 1&  &-1&  &  &-1& 1 \\ 
\AA\101\100\MM10\1100\ZZ  &   & 1&  &  &  &  & 1&  & 1&  & 1&  &  &  &  &  &  &  &  &  &-1&  & 1& 1&  &  & 1 \\
\AA10000\1\MM\100001\ZZ   &   &  &  &  &  &  &  &  &  &  &  &  &  &  &  &  &  & 1& 1&  &-1&  &  &  &  &-1&-1 \\ 
\AA\11\1001\MM1\1100\1\ZZ &   &  &  & 1&  &  & 1&  & 1&  &  &  &  &  &  &  &  &  & 1& 1& 1& 1& 1&  &  & 1&-1 \\
\AA\11000\1\MM1\10001\ZZ  &   &  &  &  &  &  &  &  &  &  &  &  &  &  &  &  &  &  & 1& 1&-1&-1&  &  &  & 1& 1 \\ 
\AA0\10001\MM01000\1\ZZ   &   &  &  & 1&  &  &  &  &  &  &  &  &  &  &  &  &  &  &  & 1&  & 1&  &  &  &-1& 1 \\
\AA0\1100\1\MM01\1001\ZZ  &   &  &  &  &  &  & 1&  &  &  &  &  &  &  &  &  &  &  &  & 1&  &-1&-1&  &  &-1&-1 \\ 
\AA00\1100\MM001\100\ZZ   &   &  &  &  &  &  & 1&  &  &  & 1&  &  &  &  &  &  &  &  &  &  &  &-1&-1&  &  & 1 \\ 
\AA000\110\MM0001\10\ZZ   &   &  &  &  &  &  &  &  &  &  & 1&  &  &  & 1&  &  &  &  &  &  &  &  &-1&-1&  &-1 \\ 
\AA0000\10\MM000010\ZZ    &   &  &  &  &  &  &  &  &  &  &  &  &  &  & 1&  &  &  &  &  &  &  &  &  &-1&  & 1 \\ 
\hline
                        &     2& 2& 2& 2& 2& 2& 2& 2& 2& 2& 2& 2& 2& 2& 2& 2& 2& 2& 2& 2&
                              $\sqrt{12}$& $\sqrt{12}$& $\sqrt{12}$& $\sqrt{12}$& $\sqrt{12}$& $\sqrt{12}$& 
                              $\sqrt{27}$  \\
\hline
\end{tabular}
\end{table}

To obtain the CGCs for the weights not listed in the tables, one can apply the 
charge conjugation operators, introduced by Moody and Patera \cite{pat_R}, 
to the dominant weight states. Charge conjugation operators $R_{\alpha_i}$, $i=1,\ldots 6$
are elements of E$_6$ and have multiple uses. The
name is derived from their property to reverse weight coordinate (``charge'') 
$w_i$ to $-w_i$. In fact, their action is up to a sign a Weyl reflection of weight $w$
into $w - w_i\alpha_i$ in the weight space. The important property for this study
is that they relate CGCs of any other weight state with those
already listed for the dominant weights. They act according to the rule
\be
R_{\alpha_i}\mid\!\! w\rangle = \exp(E_{-\alpha_i})\exp(-E_{\alpha_i})\exp(E_{-\alpha_i})\mid\!\! w\rangle .
\label{eq:R}
\ee
It is simple to check that if state $\mid\!\! w\rangle$ of the ${\bf 27}$ or ${\bf \ol{27}}$ can be lowered
by $E_{-\alpha_i}$ (see eq.(\ref{eq:e- })), $R_{\alpha_i}\mid\!\! w\rangle = E_{-\alpha_i}\mid\!\! w\rangle$.
A less trivial example for a state of the {\bf 650} is
\bea
R_{\alpha_1} \mid\!\! 2\bar{1}0001\rangle &=& 
       \exp(E_{-\alpha_1})\: \exp(-E_{\alpha_1})\: 
                  [1+E_{-\alpha_1} + (E_{-\alpha_1})^2/2]\: \mid\!\! 2\bar{1}0001\rangle =  \nnmb \\ 
   &=& \exp(E_{-\alpha_1})\: [1-E_{\alpha_1} + (E_{\alpha_1})^2/2]\: 
                  [\mid\!\! 2\bar{1}0001\rangle +\sqrt{2}\mid\!\! 000010_1\rangle + \mid\!\! \bar{2}10001\rangle] =   \nnmb \\ 
   &=& [1+E_{-\alpha_1} + (E_{-\alpha_1})^2/2]\:  \mid\!\! \bar{2}10001\rangle =   \nnmb \\ 
   &=& \mid\!\! \bar{2}10001\rangle,    
\eea
which, indeed, coincides with the reflection $(2\bar{1}0001)-2\alpha_1$ in the weight space.
The CGC decomposition of the new state is then obtained 
(using the second example in the previous section) as 
\be
R_{\alpha_1}\,[\,\mid\!\! 100000\rangle\mid\!\! 1\bar{1}0001\rangle\,] =  
      [R_{\alpha_1}\mid\!\! 100000\rangle\,]\; [R_{\alpha_1}\mid\!\! 1\bar{1}0001\rangle\,] = 
                    \: \mid\!\! \bar{1}10000\rangle\,\mid\!\! \bar{1}00001\rangle.                     \nnmb
\ee

In short, the CGCs for the {\bf 650} states are equal $+1$
if the weight does not coincide with any weight of the {\bf 78}. 
On the other hand,
the CGCs are equal to $+1/\sqrt{2}$ if the weight is other than (000000).\footnote
{
These weights (non-zero roots) can be found in \cite{AB234}, 
and also in table 20 in \cite{Slansky_PhysRep}.
}
Orthogonality of the representation spaces then implies that 
CGCs of the {\bf 78} are equal to $\pm1/\sqrt{2}$. 
Finally, CGCs for the (000000) weights are equal to
$+1/2$ for the {\bf 650}, $\pm1/\sqrt{12}$ for the {\bf 78}, and $\pm1/\sqrt{27}$ for
the singlet. Note that the signs in the decomposition of the singlet are $+$ for
even levels of the {\bf 27} and $-$ for the odd levels. (The levels of the {\bf 27} 
are listed in the last column of table \ref{t:emb}.)

\section{Application to model building: operator 27$^3$} 

Let us summarize the properties of the {\bf 27} in $E_6$ 
with respect to the SM gauge group and its branching into
$SU(3)_c\otimes U(1)_{em}$ \cite{Slansky_PhysRep}.
If we ignore the embedding where the {\bf 27} contains
a color octet, there is a unique embedding of color
with three ${\bf 3}$'s, three ${\bf \bar{3}}$'s and nine
singlets of $SU(3)_c$. 
At the level of the SM gauge group, the ${\bf 3}$'s are the 
$U$ and $D$ quark states contained in an SU(2)$_L$ doublet $Q$,
and a Higgs triplet $T$. The three ${\bf \bar{3}}$'s 
$U^c$, $D^c$, and $T^c$ are all singlets under SU(2)$_L$.
The colorless states include an SU(2)$_L$ doublet $L$, consisting of 
the left-handed neutrino and electron, two Higgs doublets $H_u$ and $H_d$ 
(two being consistent with the minimal supersymmetric extension of the SM),
and singlets $E^c$, $N^c$, and $S$.
For phenomenological reasons, the colored Higgs triplet $T$ and anti-triplet $T^c$ 
cannot enter the spectrum of the SM particles at the electroweak scale,
and, in general, are assumed to have masses close to the unification 
scale. Similarly, the SM singlet states $N^c$ and $S$ have not been
observed.

Three different embeddings of these states
into the {\bf 27} in $E_6$ are given in tables \ref{t:emb} and 
\ref{t:U1s}. They can be refered to as {\em standard embedding}, 
{\em flipped SU(5)}, and {\em flipped SO(10)}~\cite{embeddings}. 
Table \ref{t:emb} shows weights of the physical particle states 
along a subgroup chain 
\bea
E_6 &\supset& SO(10)\otimes U(1)_t \supset SU(5)\otimes U(1)_r\otimes U(1)_t \nnmb\\
    &\supset& SU(3)_c\otimes SU(2)_L\otimes U(1)_z \otimes U(1)_r \otimes U(1)_t.  
\label{eq:brnch}
\eea
The weights of SO(10), SU(5), and  SU(3)$_c\otimes$SU(2)$_L$
are obtained following the projections
\begin{equation}
\begin{array}{lcl}
(w)|_{SO(10)}&=&(w_2+w_3+w_4,w_6,w_3,w_4+w_5,w_1+w_2)|_{E_6} \\
(w)|_{SU(5)} &=&(w_1+w_2,w_3+w_5,w_4,w_2+w_3)|_{SO(10)} \\
(w)|_{SU(3)_c} &=&(w_1+w_2,w_3+w_4)|_{SU(5)} \\
(w)|_{SU(2)_L} &=&(w_2+w_3)|_{SU(5)}. 
\end{array}
\end{equation}

Table \ref{t:U1s} shows the corresponding U(1) charges. These are calculated
from \cite{Slansky_PhysRep}
\be
 Q = \bar{q}_i\, w_i
\ee
where $\bar{q}_i$'s are dual coordinates of the respective charges. In our case,
we have $\bar{q}^t = (1,-1,0,1,-1,0)$, $\bar{q}^r = (1,-1,-4,-3,-1,0)$, 
and $\bar{q}^z = (1,-1,1,-3,-1,0)$. The hypercharge, which is the U(1) factor in the SM gauge group,
is defined as 
\be
 Y \propto Q_{em} - I_3
\ee
($I_3$ is the eigenvalue of the diagonal generator in SU(2)$_L$), 
and must be contained among the three U(1) charges in (\ref{eq:brnch}).
In fact, for the first type of embedding it is equal (up to an overall factor) to  
charge $Q^z$. That makes U(1)$_z$ equivalent to the U(1)$_Y$ of the SM in this case, 
and explains the embedding's name. For the flipped SU(5) we have 
\be
\bar{q}^Y \propto 6\,\bar{q}^r - \bar{q}^z \propto (1,-1,-5,-3,-1,0)
\ee
and for the flipped SO(10)
\be
\bar{q}^Y \propto 15\, \bar{q}^t - 3\, \bar{q}^r - 2\, \bar{q}^z \propto (1,-1,1,3,-1,0).
\ee
The three embeddings in tables \ref{t:emb} and \ref{t:U1s} thus correspond to
three different embeddings of the hypercharge for the branching of $E_6$
given in (\ref{eq:brnch}). Note that other embeddings of the particle states are 
possible (we can e.g., exchange $(T^c,H_d)$ and $(D^c,L)$) but these 
correspond to the same hypercharge embedding.

\protect
\begin{table}
\caption{ 
Embeddings of the SM states into the {\bf 27} in $E_6$. 
}
\label{t:emb}
\tiny
\begin{tabular}{|ccc|cc|cc|cc|cr|}
\hline
\multicolumn{3}{|c|}{ } &\multicolumn{2}{c|}{ } &\multicolumn{2}{c|}{ } &\multicolumn{2}{c|}{ } &\multicolumn{2}{c|}{ }\\
\multicolumn{3}{|c|}{\small Superfield} & 
                                 \multicolumn{2}{c|}{\small SU(3)$_c\otimes$SU(2)$_L$ } &
                                 \multicolumn{2}{c|}{\small SU(5)} &
                                 \multicolumn{2}{c|}{\small SO(10)} &
                                 \multicolumn{2}{c|}{\small E$_6$} \\
\multicolumn{3}{|c|}{ } &\multicolumn{2}{c|}{ } &\multicolumn{2}{c|}{ } &\multicolumn{2}{c|}{ } 
                                                                                  &\multicolumn{2}{c|}{ } \\
standard & flipped & flipped &        &         &        &       &        &       &\multicolumn{2}{c|}{(100000) irrep}\\ 
embedding& SU(5)   & SO(10)  & weight &  irrep  & weight & irrep & weight & irrep & weight & level \\
\multicolumn{3}{|c|}{ } &\multicolumn{2}{c|}{ } &\multicolumn{2}{c|}{ } &\multicolumn{2}{c|}{ } &\multicolumn{2}{c|}{ }\\
\hline
 Q    &  Q    &  Q    &  (10)(1)  &  (10)(1) &  (0100)  & (0100) &  (00001)  & (00001) &  (100000)  &  0 \\
      &       &       & (\11)(1)  &          & (\1010)  &        & (\10010)  &         & (1\10010)  &  7 \\
      &       &       & (0\1)(1)  &          & (\110\1) &        & (0\1001)  &         & (10000\1)  & 11 \\
      &       &       &  (10)(\1) &          & (10\11)  &        & (010\10)  &         & (0000\11)  &  5 \\
      &       &       & (\11)(\1) &          & (0\101)  &        & (\1100\1) &         & (0\10001)  & 12 \\
      &       &       & (0\1)(\1) &          & (00\10)  &        & (000\10)  &         & (0000\10)  & 16 \\
U$^c$ & D$^c$ & T$^c$ &  (01)(0)  &  (01)(0) & (1\110)  &        & (01\110)  &         & (00\1101)  &  3 \\
      &       &       & (1\1)(0)  &          & (100\1)  &        & (10\101)  &         & (01\11\10) &  7 \\
      &       &       & (\10)(0)  &          & (0\11\1) &        & (00\110)  &         & (00\1100)  & 14 \\
E$^c$ & N$^c$ & S     &  (00)(0)  &  (00)(0) & (\11\11) &        & (\101\10) &         & (1\11\100) &  9 \\
D$^c$ & U$^c$ & D$^c$ &  (01)(0)  &  (01)(0) &  (0001)  & (0001) & (0010\1)  &         & (0\11000)  &  2 \\
      &       &       & (1\1)(0)  &          & (01\10)  &        & (1\11\10) &         & (0010\1\1) &  6 \\
      &       &       & (\10)(0)  &          & (\1000)  &        & (0\110\1) &         & (0\1100\1) & 13 \\
 L    &  L    & H$_u$ &  (00)(1)  &  (00)(1) & (1\100)  &        & (1\1010)  &         & (00010\1)  &  4 \\
      &       &       & (00)(\1)  &          & (001\1)  &        & (1000\1)  &         & (\1001\10) &  9 \\
N$^c$ & E$^c$ & N$^c$ &  (00)(0)  &  (00)(0) &  (0000)  & (0000) & (\11\101) &         & (10\1001)  & 10 \\
 T    &  T    &  T    &  (10)(0)  &  (10)(0) &  (1000)  & (1000) &  (10000)  & (10000) & (\110000)  &  1 \\
      &       &       & (\11)(0)  &          & (0\110)  &        & (0001\1)  &         & (\100010)  &  8 \\
      &       &       & (0\1)(0)  &          & (000\1)  &        & (1\1000)  &         & (\11000\1) & 12 \\
H$_u$ & H$_d$ &  L    &  (00)(1)  &  (00)(1) & (\1100)  &        & (0\1100)  &         & (001\11\1) &  5 \\
      &       &       & (00)(\1)  &          & (00\11)  &        & (001\1\1) &         & (\101\100) & 10 \\
T$^c$ & T$^c$ & U$^c$ &  (01)(0)  &  (01)(0) &  (0001)  & (0001) & (\11000)  &         & (000\111)  &  4 \\
      &       &       & (1\1)(0)  &          & (01\10)  &        & (000\11)  &         & (010\100)  &  8 \\
      &       &       & (\10)(0)  &          & (\1000)  &        & (\10000)  &         & (000\110)  & 12 \\
H$_d$ & H$_u$ & H$_d$ &  (00)(1)  &  (00)(1) & (001\1)  &        & (00\111)  &         & (01\1010)  &  6 \\
      &       &       & (00)(\1)  &          & (1\100)  &        & (01\100)  &         & (\11\1001) & 11 \\
 S    &  S    & E$^c$ &  (00)(0)  &  (00)(0) &  (0000)  & (0000) &  (00000)  & (00000) & (1\101\10) &  8 \\
\hline
\end{tabular}
\end{table}

\protect
\begin{table}
\caption{Charges for $U(1)$ subgroups of $E_6$ for the three embeddings of $U(1)_Y$. 
}
\label{t:U1s}
\tiny
\begin{tabular}{|ccc|ccc|}
\hline
\multicolumn{3}{|c|}{ } &\multicolumn{3}{c|}{ }\\
\multicolumn{3}{|c|}{\small Superfield} & {\small U(1)$_t$} & {\small U(1)$_r$} & {\small U(1)$_z$} \\
\multicolumn{3}{|c|}{ } &\multicolumn{3}{c|}{ }\\
standard & flipped & flipped &        &         &        \\
embedding& SU(5)   & SO(10)  &        &         &        \\
\multicolumn{3}{|c|}{ } &\multicolumn{3}{c|}{ }\\
\hline
 Q    &  Q    &  Q    &  1  &  1  &  1  \\
U$^c$ & D$^c$ & T$^c$ &  1  &  1  & -4  \\
E$^c$ & N$^c$ & S     &  1  &  1  &  6  \\
D$^c$ & U$^c$ & D$^c$ &  1  & -3  &  2  \\
 L    &  L    & H$_u$ &  1  & -3  & -3  \\
N$^c$ & E$^c$ & N$^c$ &  1  &  5  &  0  \\
 T    &  T    &  T    & -2  & -2  & -2  \\
H$_u$ & H$_d$ &  L    & -2  & -2  &  3  \\
T$^c$ & T$^c$ & U$^c$ & -2  &  2  &  2  \\
H$_d$ & H$_u$ & H$_d$ & -2  &  2  & -3  \\
 S    &  S    & E$^c$ &  4  &  0  &  0  \\
\hline
\end{tabular}
\end{table}

Finally, we construct a 27$^3$ operator. For 27$\otimes$27 we use the results 
of ref.\cite{pat_E6}. There is a $\ol{27}$ in this product which 
adds up with the third 27 providing for the singlet state according to the last
column of our table \ref{t:cgc000000}. Next, we show how this operator decomposes into
the SM states. 

Note that there is freedom to assign a phase to each particle state in table
\ref{t:emb}. In order to obtain the standard SU(3) and SU(2) contractions we
thus redefine the following phases 
\bea
T^c_{(01)}       & \rightarrow & - T^c_{(01)} \\
D^c_{(01)}       & \rightarrow & - D^c_{(01)} \\
U^c_{(01)}       & \rightarrow & - U^c_{(01)} \\
  D_{(\bar{1}1)} & \rightarrow & -  D_{(\bar{1}1)} \\
  Q_{(10)(1)}       & \rightarrow & -   Q_{(10)(1)} \\
  Q_{(\bar{1}1)(1)} & \rightarrow & -   Q_{(\bar{1}1)(1)} \\
  Q_{(0\bar{1})(1)} & \rightarrow & -   Q_{(0\bar{1})(1)} \\
  L                 & \rightarrow & -   L.
\eea
With these redefinitions we obtain
\bea
27^3 &=& STT^c + SH_uH_d - NLH_u + NTD^c + E^cLH_d - E^cTU^c + \\
     & & + LQT^c - U^cQH_u + D^cQH_d + D^cU^cT^c + QQT. 
\eea
This equation assumes that a cyclic
permutation is applied to the right hand side. For instance, 
$STT^c\equiv S_1T_2T^c_3 + S_2T_3T^c_1 + S_3T_1T^c_2 + 
S_1T_3T^c_2 + S_3T_2T^c_1 + S_2T_1T^c_3$.

\section{Summary}
In this paper we showed the decomposition of the 
${\bf 27}\otimes {\bf \ol{27}}$ and as a simple application we
derived the form of the 27$^3$ operator in terms of the particle
states relevant for the SM gauge group.
Since the application of ladder operators to higher irreps is derived from
their action on the lower irrep states this study represents an important first 
step in the analysis designed to provide more complete tools for $E_6$ model building.

\end{document}